\documentclass[%
 aip,
 amsmath,amssymb,
 reprint,%
]{revtex4-1}

\usepackage{graphicx}
\usepackage{dcolumn}
\usepackage{bm}

\usepackage[utf8]{inputenc}
\usepackage[T1]{fontenc}
\usepackage{etoolbox}
\usepackage{xcolor}
\usepackage{booktabs}
\usepackage{hyperref}
\usepackage{amsmath, amsthm}
\usepackage[scr=boondox]{mathalpha}

\RequirePackage{xpatch}
\xpatchcmd\citenum{\NAT@parfalse}{\NAT@partrue}{}{}

\makeatletter
\def\@email#1#2{%
 \endgroup
 \patchcmd{\titleblock@produce}
  {\frontmatter@RRAPformat}
  {\frontmatter@RRAPformat{\produce@RRAP{*#1\href{mailto:#2}{#2}}}\frontmatter@RRAPformat}
  {}{}
}%
\makeatother
\begin{document}

\preprint{AIP/123-QED}

\title[]{Discovering an interpretable mathematical expression for a full wind-turbine wake with artificial intelligence enhanced symbolic regression}
\author{Ding Wang}
\affiliation{School of Ocean and Civil Engineering, Shanghai Jiao Tong University, Shanghai, China}
\affiliation{Ningbo Institute of Digital Twin, Eastern Institute of Technology, Ningbo, China}

\author{Yuntian Chen}
\homepage{ychen@eitech.edu.cn}
\affiliation{Ningbo Institute of Digital Twin, Eastern Institute of Technology, Ningbo, China}

\author{Shiyi Chen}
\homepage{schen@eitech.edu.cn}
\affiliation{Ningbo Institute of Digital Twin, Eastern Institute of Technology, Ningbo, China}


\begin{abstract}
The rapid expansion of wind power worldwide underscores the critical significance of engineering-focused analytical wake models in both the design and operation of wind farms. These theoretically-derived analytical wake models have limited predictive capabilities, particularly in the near-wake region close to the turbine rotor, due to assumptions that do not hold. Knowledge discovery methods can bridge these gaps by extracting insights, adjusting for theoretical assumptions, and developing accurate models for physical processes. In this study, we introduce a genetic symbolic regression (SR) algorithm to discover an interpretable mathematical expression for the mean velocity deficit throughout the wake, a previously unavailable insight. By incorporating a double Gaussian distribution into the SR algorithm as domain knowledge and designing a hierarchical equation structure, the search space is reduced, thus efficiently finding a concise, physically informed, and robust wake model. The proposed mathematical expression (equation) can predict the wake velocity deficit at any location in the full-wake region with high precision and stability. The model's effectiveness and practicality are validated through experimental data and high-fidelity numerical simulations.
\end{abstract}

\maketitle

\section{Introduction}
\label{sec:intro}

\subsection{Background}
Over the past several years, there has been a surge in interest surrounding wind energy, attributed to its green, sustainable and widely accessible nature, underscoring its importance. Among the various aspects of wind power generation, understanding and modeling the wake effects of wind turbines has emerged as a crucial research area. Wake effects refer to the impact of a wind turbine on downstream flows, which can lead to reduced power product and increased aerodynamic loads of downstream turbines\cite{Port2020Wind}. Thus, investigating turbine wakes holds profound significance. To study the aerodynamic characteristics of wind turbine wakes, researchers have employed complex computational fluid dynamics (CFD) numerical methods as well as simpler engineering-focused wake models. These models, derived based on simplified conservation of physical quantities \cite{Bas}, aim to predict mean wake velocity fields and capture relevant physical characteristics. Favored in the wind industry for their simplicity and cost-effectiveness, wake models play a prominent role in predicting wake performance and effective planning for wind farms.

While several wake models have demonstrated good performance in predicting averaged characteristics in the far wake \cite{WD}, an accurate representation of the flow in the near wake, particularly near the turbine rotor, remains a challenge. The increasing density of wind farms, driven by factors like local topography, wind resources and land availability, often results in turbine spacing approaching or even falling below $3D$ (where $D$ represents the diameter of a turbine) \cite{2003Wind,Barthelmie2007}. This necessitates the development of models that can accurately capture turbine wakes both in the far wake and the near wake, even near the turbine rotor.

\begin{figure*}
    \centering
    \includegraphics[width=0.8\linewidth]{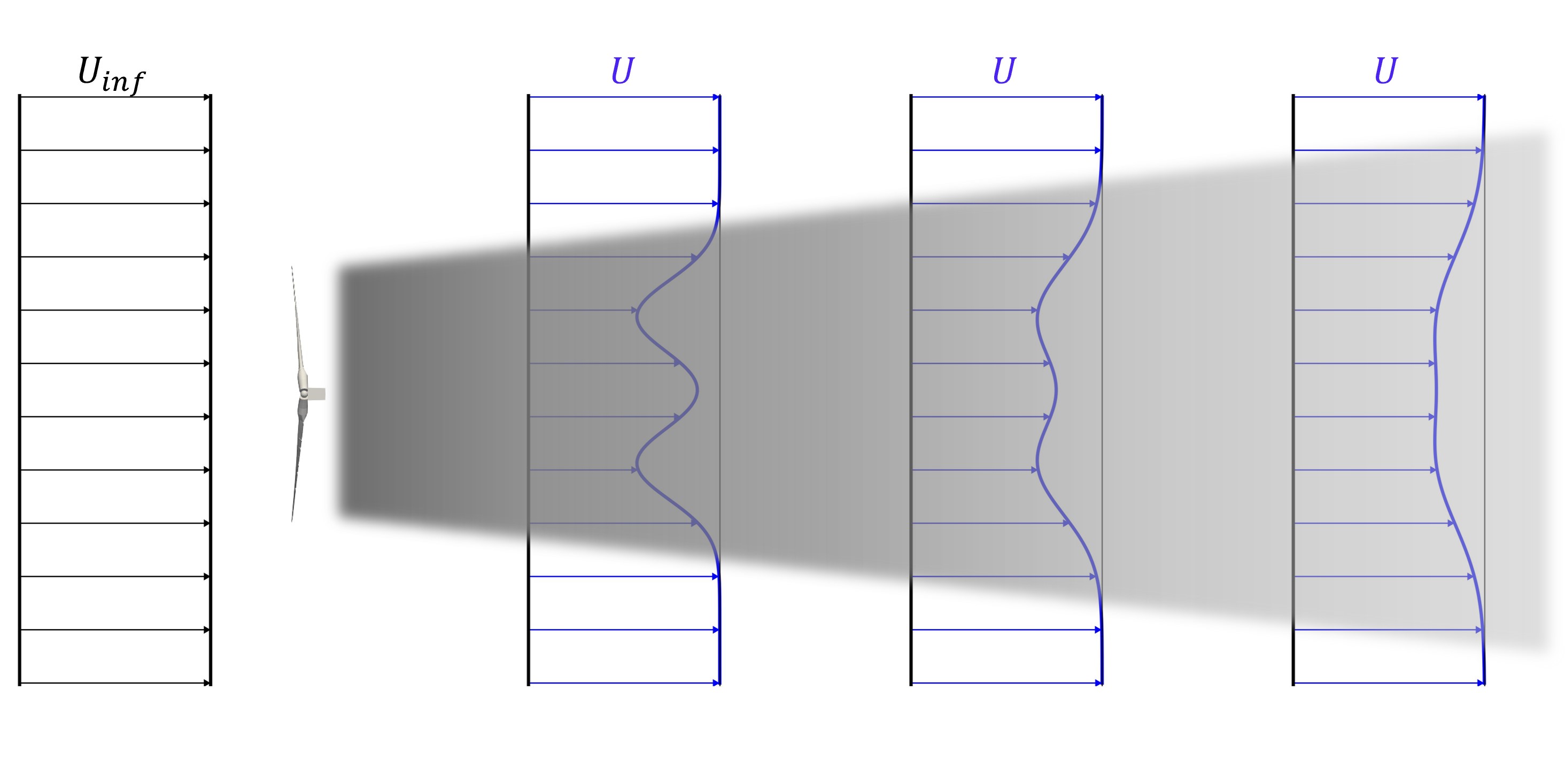}
    \caption{Schematic figure showing a wind turbine wake featuring a DG velocity deficit.}
    \label{fig:Schematic}
\end{figure*}

\subsection{Double Gaussian wake models}
According to Bastankhah \textit{et al.}~\cite{Bas}, the velocity deficit profile of wind turbine wakes can be assumed to have a self-similar Gaussian shape beyond a certain downwind distance of $x>3D$. It has been found that the velocity deficit in the near wake follows a double Gaussian (DG) distribution in wind tunnel experiments \cite{Li2016ExperimentalAN, Wang2017} and field measurements \cite{Aitken2014UtilityScaleWT}. Fig.~\ref{fig:Schematic} depicts a schematic representation of a wind turbine wake featuring a DG velocity deficit. The appearance of the two peaks (local minima) in the DG velocity distribution is believed to be caused by the minimum lift near the blade root and tip and the maximum lift in the mid of the blade \cite{Lift}. Schutz \textit{et al.}~\cite{Schutz2023} demonstrated that the DG profile is more effective in describing axial velocity distributions compared to a Gaussian profile, irrespective of whether they are in the near wake or the far wake.

By focusing on these wake characteristics, Keane \textit{et al.}~\cite{Keane2016} developed a DG wake modeling approach based on mass and momentum conservation principles. He proposed a solution that combines the velocity deficit from the wake model and the free stream, but this model still encounters challenges related to conservation. Schreiber \textit{et al.}~\cite{Schreiber2020} suggested addressing the issue of mass conservation at the outlet of the stream tube by considering an unknown downstream position instead of assuming it to be located at the turbine rotor. Keane \textit{et al.}~\cite{Keane2021} later improved his initial DG wake model by achieving global availability of the wake field through processing the modulus of complex solutions and introducing the effective radius to describe the range of wind power extraction. Furthermore, Gao \textit{et al.}~\cite{Gao2022} introduced a DG model that incorporates the influence of vertical wind shear, and observed higher accuracy when considering delay time. Soesanto \textit{et al.}~\cite{Soesanto2022} investigated the anisotropy of wake expansion and defined the onset of the far wake region with a scaling factor of utility-scale wind turbines based on the model proposed by Keane \textit{et al.}~\cite{Keane2021}. In subsequent work, Soesanto \textit{et al.}~\cite{Soesanto2023} examined the predictive performance of the newly proposed DG model under different turbulence intensities. Additionally, Lanzilao \textit{et al.}~\cite{wakeMerge} introduced a new method for wake merging to study wind farms, which deals with various models including the DG model.

However, the DG models derived from theoretical derivations still have unresolved issues that require attention. The primary challenge lies in the inability to handle the wake flow near the turbine rotor, mainly due to physical conservation principles. These methods rely on simplified momentum conservation equations, neglecting factors such as the pressure gradient and viscosity. Furthermore, they only consider streamwise velocity and apply these assumptions to a hypothesized velocity deficit distribution containing amplitude functions. As a result, these models may not provide genuine physical explanations in areas where assumptions break down. Inconsistencies also arise in the hypotheses about wake characteristics, including the wake expansion ($\sim x^{1/3}$ \cite{Keane2016}, $\sim x^{0.75}$ \cite{Keane2021}, $\sim x$ \cite{Soesanto2022}) and wake radius ($3\sigma$ \cite{Keane2021}, $2.81\sigma$ \cite{Gao2022}, $2.58\sigma$ \cite{Soesanto2023}, where $\sigma$ is the standard deviation), among others. These variations in definitions may significantly impact the performance of these theoretically derived models under specific circumstances. Finally, simply fitting quite a few parameters that need to be given in DG wake models by constants introduces errors. Consequently, deriving DG models from first principles via physically driven methodologies remains exceedingly challenging, as the aforementioned issues persist unresolved.

\subsection{Knowledge discovery methods}
The complicated aerodynamic characteristics of wind turbine wakes present an ideal environment for the application of data-driven techniques in wake modeling, capable of accommodating high-dimensional nonlinear mapping relationships. However, achieving desired performance directly through machine learning is challenging. While neural networks theoretically possess universal approximation capabilities, they lack explicit knowledge \cite{KD}. This "black box" nature affects practicality by hindering a deeper comprehension of underlying physical processes. Hence, there exists a necessity to extract interpretable knowledge from the data using interpretable machine learning methods. The essence of wake modeling lies in elucidating what wind turbine wakes look like through simple expressions, constituting a process of knowledge discovery. Thus for these expressions, a competent knowledge discovery approach should produce an interpretable, concise and precise model \cite{DL-PDE}. SR, as a data-driven approach to uncovering potential physical laws and control equations from data, significantly advances modeling, simulation, and understanding across scientific and engineering domains. For instance, Zeng \textit{et al.}~\cite{Zeng2023DeepLD} demonstrated the superior performance of partial differential equations (PDEs) discovered through a knowledge discovery method over theoretically derived ones in predicting the evolution of viscous gravity currents. Similarly, Cranmer \textit{et al.}~\cite{Symbolic-Distillation} unveiled an explainable law of dark matter. Leveraging knowledge discovery methods to unveil inherent laws latent in data may present a novel approach to wake modeling.

SR, a method of supervised learning, deduces symbolic mathematical expressions from a given dataset. In such algorithm, precision and interpretability are pursued within a domain of straightforward analytical expressions \cite{PYSR}. Presently, various approaches to SR have been proposed, generally incorporating regression-based and expression tree-based \cite{Makke2022}. Linear regression-based methods operate under the assumption that the target symbolic expression linearly combines nonlinear functions of the operands in the candidate library. This simplifies the SR task to resolving a system of linear equations. SINDY \cite{SINDY} stands out among such methods, particularly in discovering PDEs \cite{PDE-FIND}, although constrained by a predefined candidate library. Consequently, Xu \textit{et al.}~\cite{DLGA-PDE} and Chen \textit{et al.}~\cite{SGA} expanded the solution space by respectively introducing the genetic algorithm and binary trees. EQL \cite{EQL}, a nonlinear regression-based method, utilizes symbolic mathematical operations to seek target expressions instead of activation functions in artificial neural networks. In fact, the prevalent approach to visualizing symbolic expressions involves employing the tree topology comprising branches and nodes \cite{Angelis2023ArtificialII}. In this method, mathematical expressions are represented as binary trees composed of internal nodes and terminal node (leaves). Each internal node denotes an operator from the candidate library (e.g. $+, -, \times, \div, \sin{}, \exp{}$), while each terminal node signifies an operand, namely, an input variable or constant. Ultimately, the components of these trees are integrated to generate expressions. Recent years have witnessed numerous innovative developments in SR, with a primary focus on genetic algorithms applied to tree-based methods \cite{PYSR}. Results from the SRBench competition reveal that contemporary SR methods based on genetic algorithms outperform new methods rooted in other optimization domains in ground-truth problems and real-world problems \cite{SRBench,Frana2023InterpretableSR}. Genetic algorithms, inspired by processes of natural evolution, offer a method for discovering the fittest solutions. These algorithms mimic natural selection, heredity and mutation to optimize the candidate expressions in a population iteratively \cite{Physo}. Additionally, various studies have proposed different implementations of SR, including reinforcement learning \cite{DSR,RDISCOVER,DISCOVER}, Markov chain Monte Carlo-like sampling \cite{BSR}, and transformer-based methods \cite{SR-transformer,NSRwH}.

\subsection{Knowledge embedding methods}
Knowledge discovery and knowledge embedding are believed to serve as vital bridges connecting data and knowledge. Purely data-driven methods solely rely on data, necessitating substantial amounts, and may yield models that contradict physical mechanisms. In practical applications, incorporating domain knowledge into models offers more comprehensive information, enhancing the efficiency, robustness and accuracy of data-driven approaches \cite{KD}. The importance of knowledge embedding is further underscored by the success of physics-informed neural networks (PINNs) that incorporate physical constraints into the loss function \cite{PINN,R-DLGA,NC_Sun}. SR benefits from identifying shared traits and attributes present in physical systems, which can effectively reduce the search space and accelerate to find the optimal solution. These methods often utilize various characteristics such as symmetry \cite{AIFA,Reinbold2021RobustLF,NSRwH}, parity\cite{Barber2021PhysicalCE}, separability\cite{Symbolic-Distillation} and dimensional conservation\cite{ZangTAML,Physo}. In summary, researchers are supposed consider mature physical expressions (analytic or empirical), and make appropriate modifications based on the specific problems under investigation \cite{Medina2023ActiveLI,Angelis2023ArtificialII}. 

Therefore, a salient question pertains to the identification of models that precisely describe the mean flow throughout the wind turbine wake, encompassing both the near and far wake regions. This study employs SR as a data-driven method to extract an interpretable expression from simulation data. Our objective is to integrate domain knowledge concerning the DG velocity deficit in wind turbine wakes into SR, thereby facilitating knowledge discovery for modeling a full wind-turbine wake. The modeling of wind turbine wakes presents a significant challenge for SR within a real-world context. In contrast to existing SR targets, which predominantly focus on confirming known equations, our endeavor seeks to unveil an expression describing the mean flow within wind turbine wakes, a previously unavailable insight. This study epitomizes a practical application of knowledge discovery, contributing substantively to the expansion of human cognitive frontiers. 

This paper is organized as follows. In Sec.~\ref{sec:method}, we first describe the training data of the wind-turbine wake velocity field and the domain knowledge of the DG profile of the velocity deficit. Then, we introduce the SR method and the criteria for selecting expressions. In Sec.~\ref{sec:results}, we present the expression discovered by SR for predicting the wind turbine wake. The validation of the experimental data and high-fidelity CFD data is conducted in Sec.~\ref{sec:validation}. Finally, we draw conclusions in Sec.~\ref{sec:conclu}.

\section{Methodology}
\label{sec:method}

\subsection{Data preparation}
\label{subsec:data}
The recent rapid advancements in computer hardware and numerical algorithms have propelled numerical simulation methods to the forefront of wind turbine performance optimization. In support of the data-driven approach, we utilize a CFD method, which aids in obtaining expensive flow parameter estimates that are otherwise difficult to acquire through experimental means. Large-eddy simulations (LES) represent an efficient CFD tool for investigating wind farms, offering detailed insights into the aerodynamics of wind turbine wakes. Utilizing the numerical solver "SOWFA" \cite{SOWFA}, we perform LES of a wind turbine operating in a fully-developed turbulent atmospheric boundary layer (ABL). The simulation process consists of two stages: the precursor simulation and the turbine simulation.

During the precursor stage, turbulent ABL flow over terrain representative of level grassland under neutral stability is simulated. The computational domain, measuring $1800\times1000\times600$~m$^3$, is discretized into a grid resolution of $180\times100\times80$ in the streamwise ($x$), spanwise ($y$) and vertical ($z$) directions. Periodic boundary conditions are applied in the horizontal directions, with a wall model is employed at the bottom boundary. To alleviate the mismatch in the logarithmic layer, the velocity at the fourth grid point is matched with the wall shear stress \cite{2012Wall}. We use the Lagrangian-averaged scale-dependent dynamic model \cite{2005LASD} to model the subgrid-scale effects, which accounts for the anisotropy of eddy viscosity and the scale dependence of the Smagorinsky coefficient. Upon reaching a statistically steady state, velocity fields at the inflow boundary are saved as the inlet condition for subsequent simulations.

In the turbine simulation stage, we simulate a single wind turbine operating in the turbulent ABL obtained from the precursor simulation. The turbine considered here is the NREL 5MW reference HAWT, featuring a rotor diameter ($D$) of $126$~m and a hub height of $90$~m. The turbine effect is modeled by incorporating the volumetric force into the momentum equation using the actuator disk method with rotation. With the turbulence intensity at hub height of $4.7\%$, the turbine rotates at a speed of $9$ rotations per minute, corresponding to the optimal tip-speed ratio for this particular turbine model. These computational settings are consistent with previous studies, and further computational details and validations of the LES solver can be found in Ref.~[\citenum{DC,WD,FF}].


\subsection{Domain knowledge in wind turbine wakes}
\label{subsec:domain_knowledge}
\begin{figure*}
    \centering
    \includegraphics[width=0.6\linewidth]{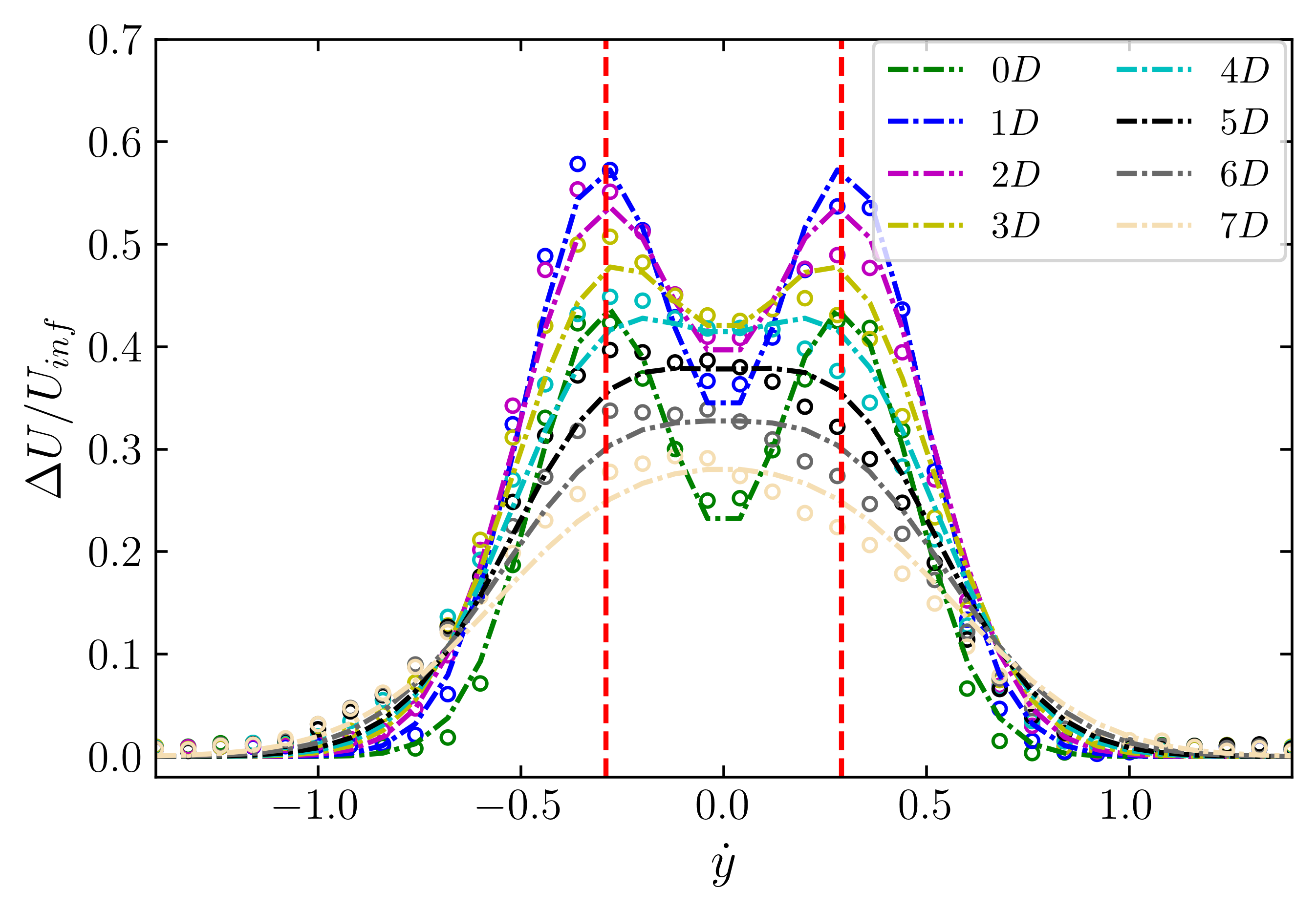}
    \caption{Profiles of the velocity deficit at different downstream locations. Circles represent LES results and dashdotted lines represent DG fittings. Vertical red dashed lines are at $\dot{y}=\pm0.29$.}
    \label{fig:similarity}
\end{figure*}
Fig.~\ref{fig:similarity} illustrates the velocity deficit at different downstream locations in the wind turbine wake obtained from the CFD simulations. It is observed that the velocity deficit across the entire wake field can be precisely characterized using a DG fit. The DG fit successfully captures the local maximum of the velocity deficit near the hub ($\dot{y}=\pm0.29$) and the local minimum at the wake center in the near wake. Therefore, the velocity deficit in the wind turbine wake is represented by the following DG function:
\begin{equation}
\frac{\Delta U(x,y)}{U_{inf}} = a \times \left(\exp{(-\frac{(\dot{y} -\mu)^2}{2\sigma^2})}+\exp{(-\frac{(\dot{y}+\mu)^2}{2\sigma^2})} \right),
\label{eq:DG}
\end{equation}
where $\dot{x}=x/D$ and $\dot{y}=y/D$ represent the normalized positions in the flow field, $\Delta U/U_{inf}$ denotes the normalized velocity deficit. The parameters amplitude $a$, mean $\mu$ and standard deviation $\sigma$ depend solely on the streamwise position $x$, i.e. $a=a(\dot{x})$, $\mu=\mu(\dot{x})$, $\sigma=\sigma(\dot{x})$. The DG function describes the distribution of wake velocity in the spanwise direction, while changes in the streamwise direction are captured by the parameters amplitude $a$, mean $\mu$ and standard deviation $\sigma$.

Modeling wakes using data-driven approaches poses a challenge due to the limited data available. In such small-sample scenarios, leveraging domain knowledge and expert experience to assist analysis and prediction can effectively enhance the capabilities of knowledge discovery methods for data mining and pattern recognition \cite{NRP,KD}. This allows for deeper insights to be extracted from the limited data. 

In this study, the premise that the wake velovcity deficit conforms to a DG distribution is incorporated as prior knowledge within SR. Following a similar rationale to the approach outlined in the study of Lou \textit{et al.}~\cite{lou2024empowering}, we adopt a hierarchical equation structure informed by domain knowledge due to the scarcity of available data. Subsequently, SR is employed to discern the constituent sub-models within this structure. Specifically, we decompose the complex wake velocity deficit field $\mathcal{G}(x,y)=\mathcal{F}(\mathcal{H}(x),y)$ into two sub-models $\mathcal{F}$ and $\mathcal{H}$, where the wake streamwise variation $\mathcal{H}(x)$ and the spanwise direction $y$ are considered mutually independent. Furthermore, the functional form of $\mathcal{F}$ is pre-determined to be DG (Eq.~\ref{eq:DG}), so our task is to find the function $\mathcal{H}$ with respect to $x$. This approach achieves dimensionality reduction in physics and feature selection in SR. The essence of genetic SR algorithms is a brute-force process with an exponentially increasing number of the input features and operators. This makes the computational cost required for evaluating tens of thousands or more expressions a significant drawback of SR. Therefore, in our wake modeling study, SR is performed simultaneously for the parameters amplitude $a$, mean $\mu$ and standard deviation $\sigma$, with the corresponding input features (variables) all being the normalized streamwise position $\dot{x}$. Leveraging this prior domain knowledge is beneficial in defining the candidate library, leading to a reduction in the search space. Similarly, Chen \textit{et al.}~\cite{TAML} demonstrated the feasibility of using indirect supervised learning with intermediate variables by integrating information on physical mechanisms with data, while Cranmer \textit{et al.}~\cite{Symbolic-Distillation} and Tenachi \textit{et al.}~\cite{Physo} successfully utilized SR to recover the correct expressions through the separability of data.

\subsection{Symbolic regression}
\label{subsec:sr}
SR is a technique employed to discover expressions approximating datasets or the outputs of arbitrary functions. It involves searching through a set of variables, operators and constants that can be selected and combined, with evaluations based on both accuracy and simplicity. In this study, we utilized the Julia library "PySR"\cite{PYSR} for SR. PySR employs an evolutionary algorithm with multiple populations to search for symbolic expressions optimized for the loss function. It converges solutions by optimizing the symbolic expressions in populations through continuous tournament selections and mutations. In the SRBench competition at the 2022 GECCO conference, PySR demonstrates excellent performance in expression simplicity, feature selection and extrapolation accuracy\cite{GECCO}.

\begin{figure*}
    \centering
    \includegraphics[width=0.65\linewidth]{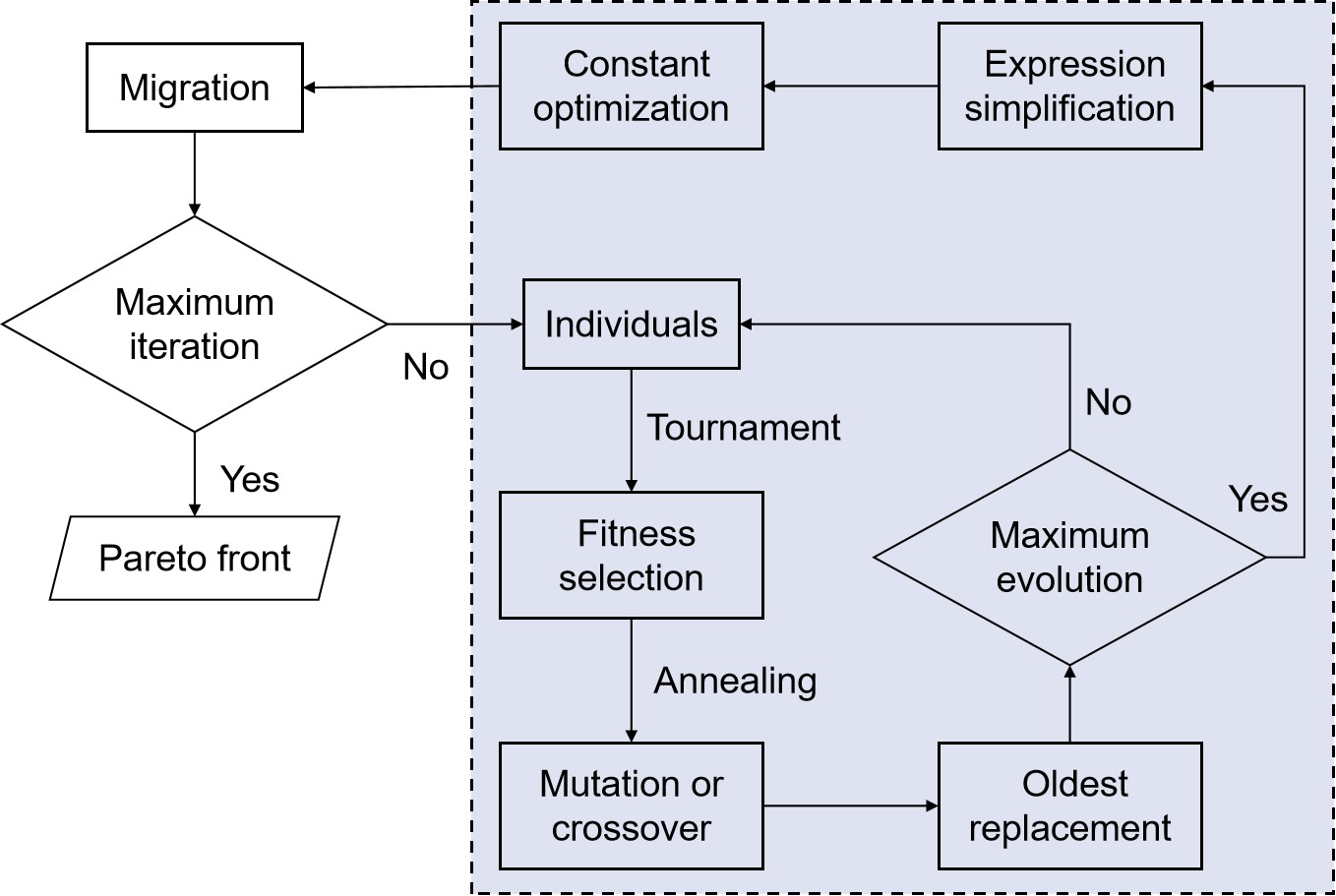}
    \caption{The workflow scheme of SR. The blue box represents operations within a population.}
    \label{fig:Flow chart}
\end{figure*}

Variables, operators and constants are stored as nodes within binary trees. These trees represent individual expressions, which collectively form a population. Evolutionary methods are then applied across multiple populations to select the fittest individuals based on predefined criteria, discussed further in Sec.~\ref{subsec:Fitness}. Tournament selection \cite{tournament} is employed asynchronously by each population to select individuals. Unlike standard operations, this selection process prioritizes longer-standing individuals over less fit ones. Inter-population migration helps mitigate the risk of getting trapped in local optima. This Genetic algorithm encompass mutation and crossover, offering eight mutation choices such as operator mutation and node insertion, detailed in Appendix~\ref{app:evol}. Simulated annealing \cite{annealing} is integrated into the optimization strategy to maintain species diversity during the evolutionary process. Following individual evolution, expressions are further simplified and constant optimization are carried out using the BFGS algorithm \cite{BFGS}. The workflow of SR is depicted in Fig.~\ref{fig:Flow chart}.

\begin{figure*}
    \centering
    \includegraphics[width=0.85\linewidth]{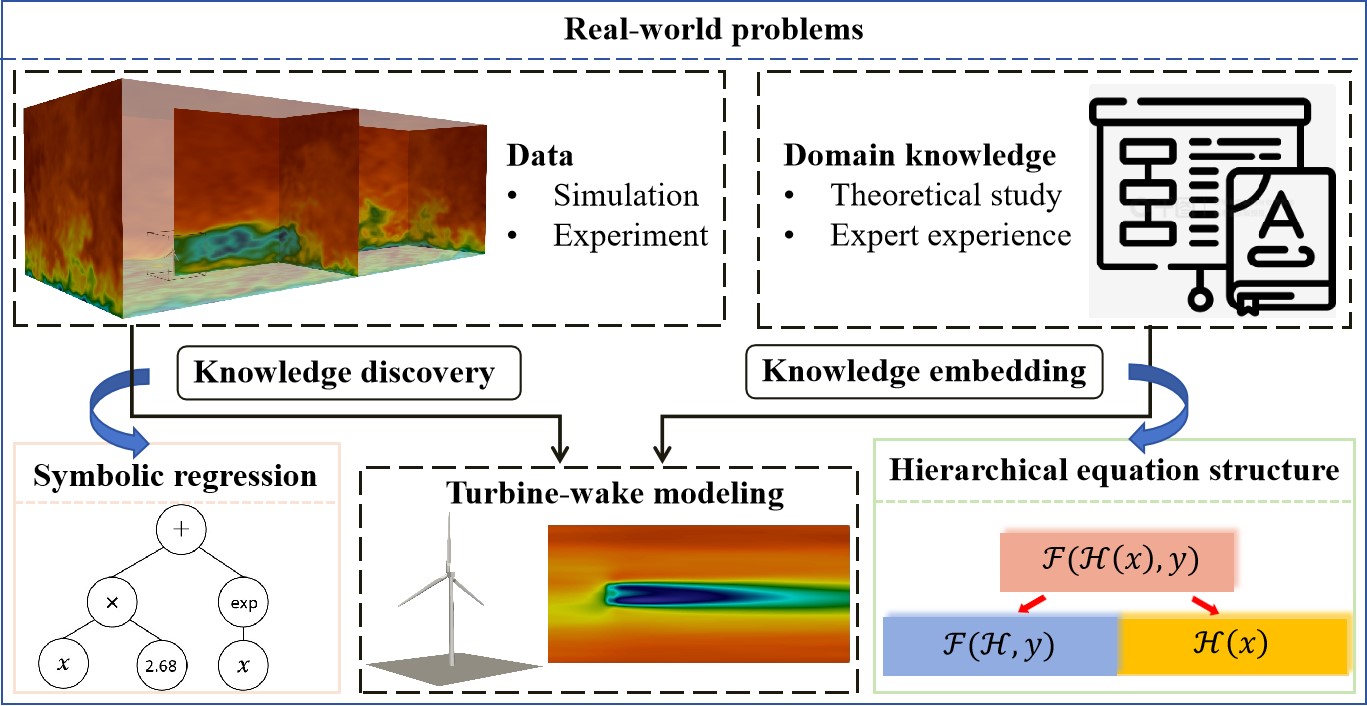}
    \caption{The Schematic diagram of domain knowledge-driven genetic SR framework for the turbine-wake modeling, a real-world problem.}
    \label{fig:framework}
\end{figure*}

The input variable is the normalized streamwise position $\dot{x}$ in CFD data, with a symbolic candidate library "$+$,$-$,$\times$, $\div$, $\exp$, $\cos$, $\wedge$". The dataset for DG fitting consists of $125 \times 36$ data points, encompassing both streamwise and spanwise directions. Consequently, the dataset size for SR stands at $125 \times 3$, focusing on three parameters across $125$ streamwise positions. Imposing constraints on operators helps reduce the search space and yields more interpretable expressions. We therefore restrict the right argument of the binary operator $\wedge$ to only be a leaf node, allowing expressions like $x^{x}$ while disallowing $x^{x+1}$. Additionally, nesting of $\exp()$ within $\exp()$ and that of $\cos()$ within $\cos()$ are prohibited, such as $\exp{(\exp{x})}$. SR is conducted simultaneously on parameters amplitude $a$, mean $\mu$ and standard deviation $\sigma$, with 20,000 iterations resulting in three expressions. Thus, our study involves two stages: (i) conducting CFD simulations to obtain wind-turbine wake data, fitting velocity deficit with parameters $a$, $\mu$ and $\sigma$ using a DG function, and (ii) performing SR to output corresponding expressions, as shown in Fig.~\ref{fig:framework}.

\subsection{Fitness and score}
\label{subsec:Fitness}
Fitness evaluates the performance of each individual in addressing the target problem, guiding the selection procedure to favor fitter individuals for contributing genetic material to subsequent generations, thus facilitating the evolution of better solutions over time \cite{Medina2023ActiveLI}. Increasing the number of equation terms can indeed enhance fitting accuracy; however, an excess of terms may lack clear physical significance. Consequently, SR prioritizes a profound comprehension of the physical mechanisms over precise data fitting, acknowledging the potential influence of errors in measurements and computations \cite{NSRwH}. The fitness criterion therefore combines accuracy, determined by mean square error (MSE) and simplicity, reflected in the complexity. The complexity of an expression is computed by summing the complexity of all nodes in the expression tree. We uniformly define the complexity of each node, making an individual's complexity $c$ equivalent to the number of nodes. The fitness is then expressed as follows:
\begin{equation}
F=\frac{\sum_i^N \left(\mathscr{h}(x_i)-\mathscr{Y_i} \right)^2}{N} \times \exp{(f_c)},
\label{eq:fitness}
\end{equation}
where $\mathscr{h}(x)$ represents the expression discovered by SR about one of the three target problems, $\mathscr{Y}$ represents the true values correspondingly, and $N$ is the number of data points. $f_c$ is the frequency of the expression complexity in the population considering recency, calculated by the number of expressions at this complexity with a moving window, whereby ones appearing early are excluded from the count. This ensures roughly equal numbers of expressions at each complexity, encouraging exploration in evolution. Ultimately, the optimal expressions at different complexities discovered from the infinite set of potential solutions form the Pareto front.

In this study, we perform SR on three parameters (amplitude $a$, mean $\mu$ and standard deviation $\sigma$) simultaneously, resulting in three sets of Pareto fronts. These parameters are then used in Eq.~\ref{eq:DG} to generate the final expression $\mathcal{F}(\mathcal{H}(x),y)$ that describes the wake velocity deficit. Therefore, the final loss is calculated as $L=\lVert \mathcal{F}(\mathcal{H}(x),y)-\mathbb{Y} \rVert^2_2 /N$, where $\mathbb{Y}$ represents the true values of the normalized wake velocity deficit, and the final complexity is determined by $C=c_{a}+c_{\mu}+c_{\sigma}$. The combination of these three Pareto fronts are sorted by complexity $C$ to form a new Pareto front $\mathbb{P}$.

Modeling turbine wakes presents a real-world problem distinct from a ground-truth problem with the standard answer. Consequently, an additional criterion is required to determine the preferred analytic expression among the optimal expressions with varying complexities in $\mathbb{P}$ at the conclusion of SR. The objective of knowledge discovery is to achieve the optimal balance between expression accuracy and simplicity \cite{KD}. While an open search space facilitates the discovery of approximately equivalent equations, it often yields overly complex expressions with better accuracy, hindering the extraction of meaningful physical insights \cite{SRBench}. Thus, we aim to identify the optimal expression that attains regression quality while ensuring simplicity, thereby enhancing interpretability and generalization ability. To attain this goal, a scoring system is utilized as a filtering criterion:
\begin{equation}
    S_i=-\frac{\log(L_i/L_{i-1})}{C_i-C_{i-1}},
\label{eq:score}
\end{equation}
where $i$ represents the position of the expression in $\mathbb{P}$. The purpose of scoring is to quantitatively assess whether increasing complexity to reduce error is worthwhile. In simpler terms, if a slight increase in complexity $C$ results in a substantial reduction in loss $L$, then that specific expression should be chosen. To ensure simplicity, the maximum complexity for the three parameters in SR is set as $c_{a,max}=c_{\mu,max}=c_{\sigma,max}=20$. Finally, as the threshold, we set $1.2$ times the minimum loss in $\mathbb{P}$, and select the expression with the highest score within this range as the final result.

\section{Results and discussion}
\label{sec:results}

\subsection{Parameters for amplitude, mean and standard deviation}
\begin{figure}
    \centering
    \includegraphics[width=1\linewidth]{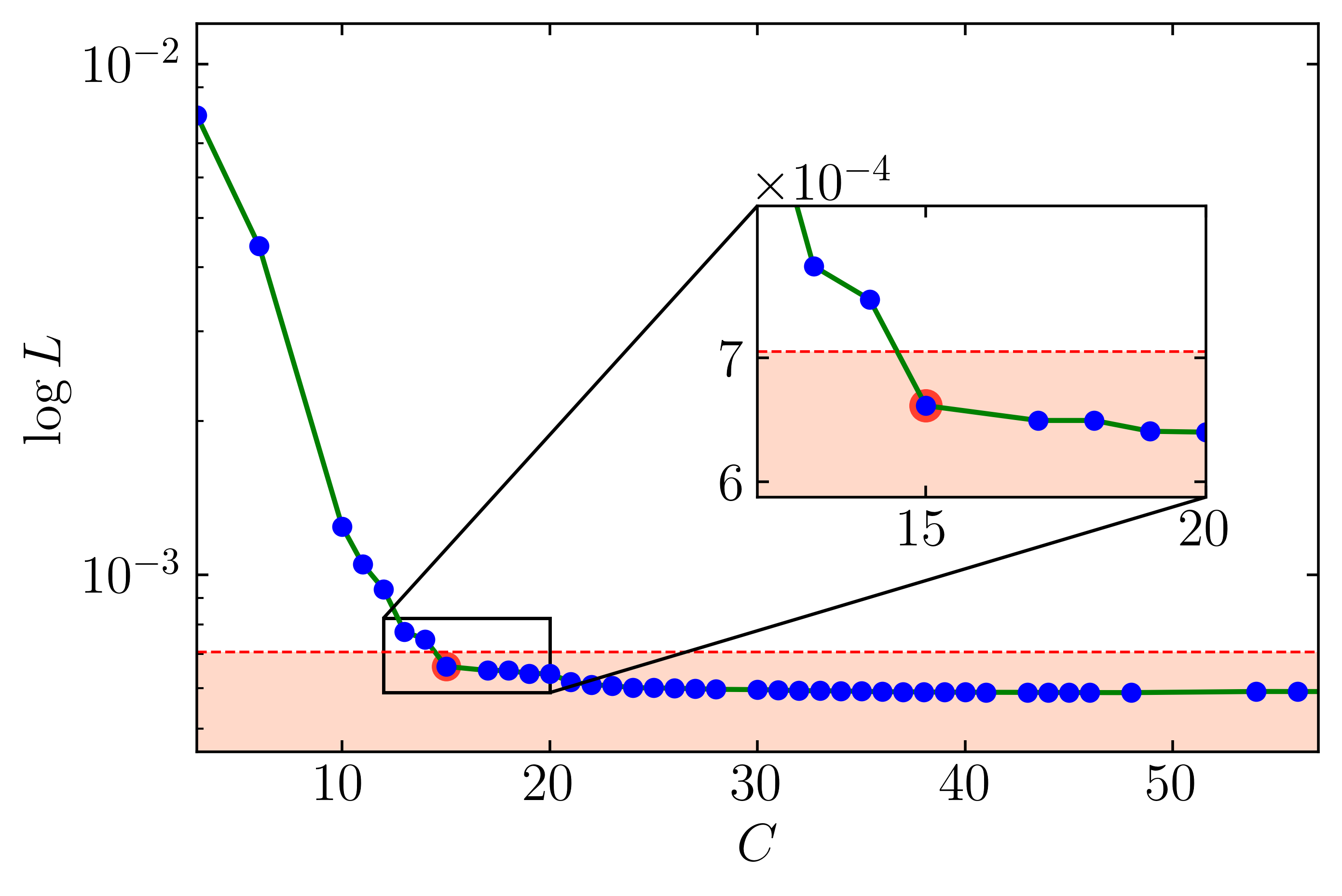}
    \caption{The relationship between the logarithmic loss and the complexity of the expressions obtained in $\mathbb{P}$. The green line denotes the Pareto curve, which consists of blue points representing the expressions in $\mathbb{P}$. The red dashed line represents the threshold, and the red point corresponds to the optimal expression.}
    \label{fig:curve}
\end{figure}

The expressions discovered through SR for the parameters amplitude $a$ mean $\mu$ and standard deviation $\sigma$ are respectively listed in Appendix~\ref{app:exp}. The complexity and the loss of the final expression for the wake velocity deficit are presented in the form of a Pareto curve in Fig.~\ref{fig:curve}. The score is essentially the negative value of the slope of the curve. Within the selected range of loss, the optimal expression with the fastest increase in accuracy as complexity increases is chosen based on the scoring mechanism.

\begin{figure}
    \centering
    \includegraphics[width=\linewidth]{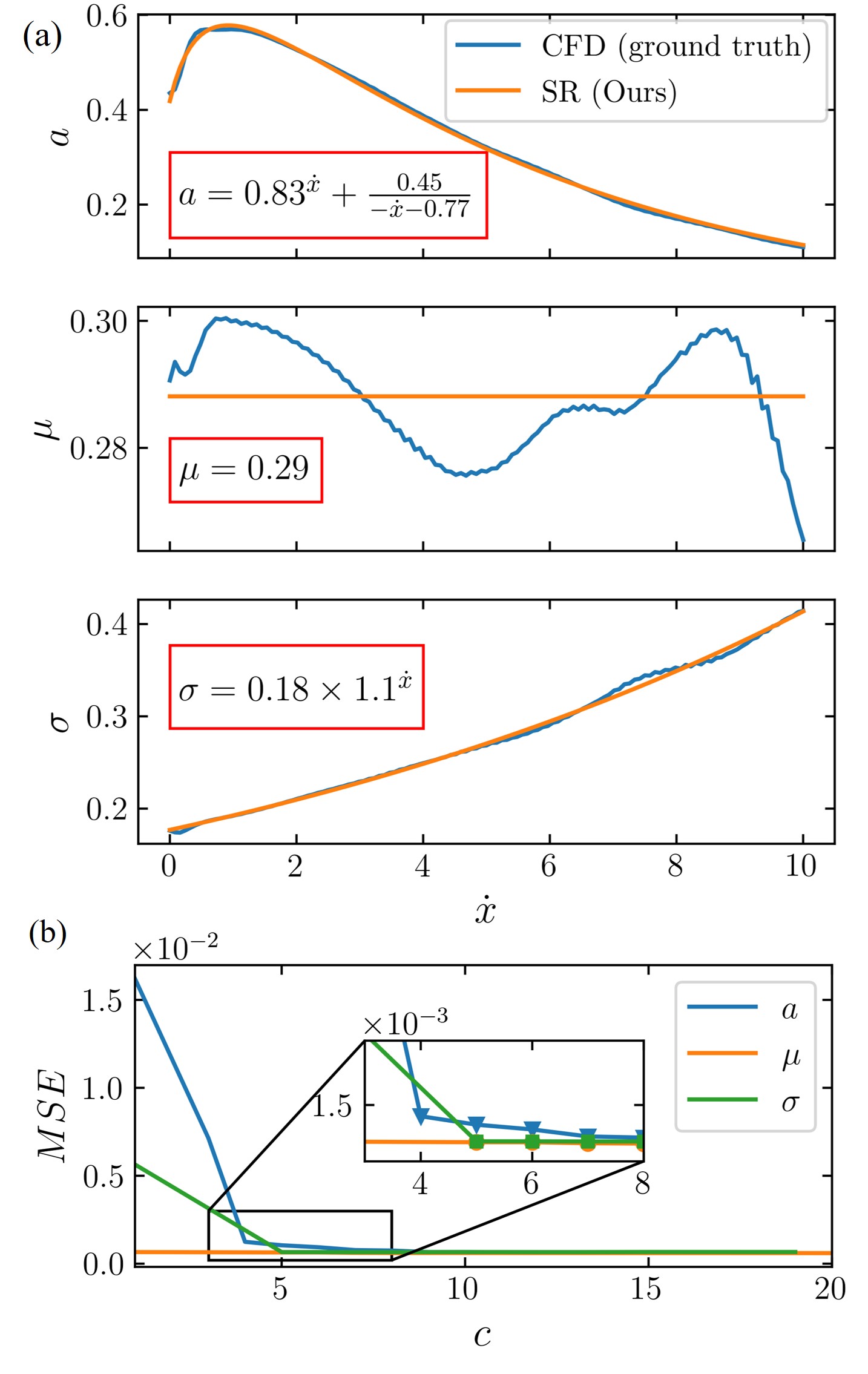}
    \caption{(a) Expressions of parameters amplitude $a$, mean $\mu$ and standard deviation $\sigma$ with respect to the streamwise distance $x$ discovered through SR. (b) Errors for expressions with different complexity regarding to the three parameters.}
    \label{fig:3 parameters}
\end{figure}

We present the specific results obtained from the application of SR to determine the best expressions for the three parameters, as shown in Fig.~\ref{fig:3 parameters}. These parameters are associated with the wake performance downstream the wind turbine. Peaks in the DG profile represent the maximum velocity deficit at each downstream position. It should be noted that the peak of one Gaussian in the DG function here is little affected by the other. Therefore, the amplitude $a$ is closely related to the DG peaks, specifically $\Delta U(\dot{x},\dot{y}=\pm\mu)/U_{inf} \approx a(\dot{x})$. We have found that SR accurately captures the maximum velocity deficit throughout the wake region at $x\approx 1D$. The second term in the amplitude expression characterizes the velocity deficit's increasing trend near the turbine rotor ($x<1D$) in fractional form. This behavior, showing an initial increase followed by a decrease, has been previously observed in wind tunnel experiments and high-fidelity CFD results \cite{Hancock2014,DC}. In contrast, traditional wake models, based on simplified conservation equations and assumptions, assume a monotonic decrease in the maximum deficit downstream. 

Regarding the mean $\mu$, we have chosen to utilize a simple expression with a complexity of one ($c=1$), represented by a constant, in order to reduce the overall complexity of the model, while ensuring high accuracy. This choice aligns with the approach taken by other researchers in previous studies on DG wake models, where constants are used to approximate or fit the mean $\mu$ \cite{Schreiber2020,Gao2022,Soesanto2023}. Fig.~\ref{fig:3 parameters}(b) illustrates the errors in the final prediction of the wake velocity deficit for expressions of the three parameters varying in complexity. The remaining two parameters are held at their optimal choices solely to evaluate the performance of individual parameter selections. Since the variation magnitude of the mean is orders of magnitude smaller than the other two parameters, employing this simplified representation has negligible impact on the final model. Additionally, Keane \textit{et al.}~\cite{Keane2021} suggests that this parameter may be dependent on the aerodynamics of a specific wind turbine.

In Gaussian-like wake models, the standard deviation $\sigma$ can effectively capture the wake expansion. In contrast to the commonly assumed linear wake expansion, the discovered expression reflects an increasing wake expansion rate in the form of a power function. This can be attributed to mixing between wind turbine wakes and ambient turbulence, which significantly affects wake recovery, particularly at farther downstream positions ($x>8D$).

\begin{figure}
    \centering
    \includegraphics[width=\linewidth]{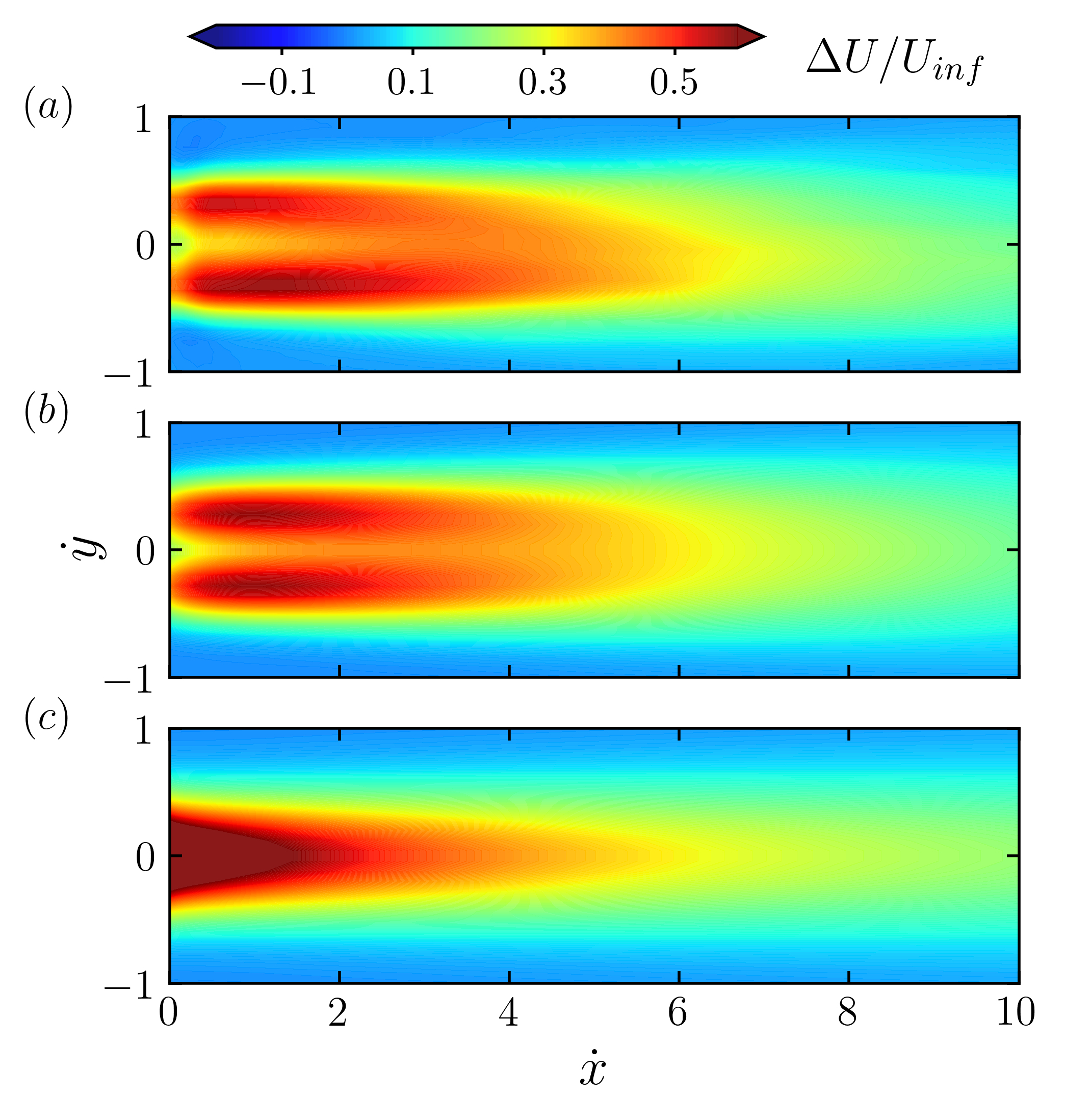}
    \caption{Contours of the velocity deficit from (a) CFD, (b) the SR expression and (c) the Bastankhah wake model.}
    \label{fig:contour_wakeDeficit}
\end{figure}

\begin{figure*}
    \centering
    \includegraphics[width=0.65\linewidth]{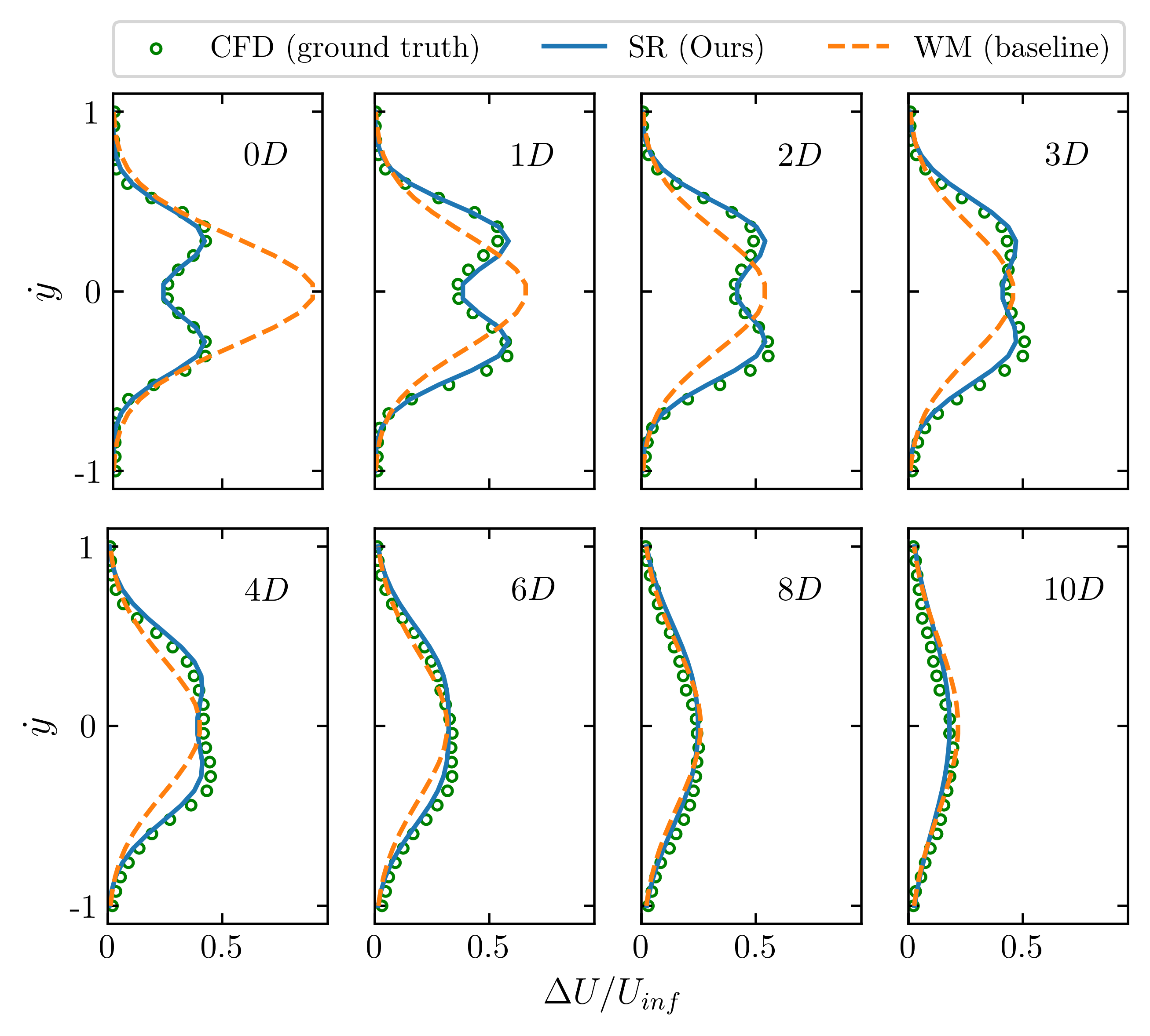}
    \caption{Comparison of velocity deficit profiles from CFD, the SR expression and the Bastankhah wake model at several stations.}
    \label{fig:8}
\end{figure*}

\subsection{The wake velocity deficit}
In Fig.~\ref{fig:contour_wakeDeficit}, the results of wind turbine wakes from CFD, SR, and the Bastankhah wake model (WM) are compared. The Bastankhah wake model \cite{Bas} is a widely used Gaussian wake model, with more details provided in the Appendix~\ref{app:Bas}. Furthermore, Fig.~\ref{fig:8} illustrates the velocity deficit at eight different downstream locations, including both near-wake and far-wake regions, obtained from the three methods. It can be observed that the Bastankhah wake model tends to fail in accurately computing the velocity deficit in the near wake. This model, designed for the far wake, significantly overestimates the velocity deficit near the wake center, particularly around the turbine rotor $(x<1D)$. This result arises because the flow conditions in this region do not satisfy the assumptions of the wake model, which is based on no pressure gradient, no viscosity and only axial velocity. On the other hand, the model discovered by SR accurately predicts the velocity deficit in the near wake, encompassing both the wake edge and the wake center. Although the Bastankhah wake model has been relatively successful in dealing with the velocity deficit in the far wake, the expression discovered exhibits higher accuracy. Wake models employing linear expansion tend to underestimate the wake width in the farther downstream far-wake region $(x>8D)$, leading to a slight overestimation of the wake deficit. We found that, compared to the Gaussian wake model, the DG model discovered by SR can accurately capture the variation of wake velocity deficit with downstream position, increasing before decreasing, as indicated by the amplitude $\alpha$ expression. Data-driven methods can capture this phenomenon, whereas wake models only reflect their fundamental assumption: wake expansion is linear in all wake regions. It is worth noting that wind turbine wakes from CFD are not entirely symmetric in the spanwise direction, primarily due to the rotation of the wind turbine. For practical applications, engineering-focused wake models typically assume spanwise symmetry, and thus, our method also maintains this symmetry.

\begin{figure*}
    \centering
    \includegraphics[width=0.85\linewidth]{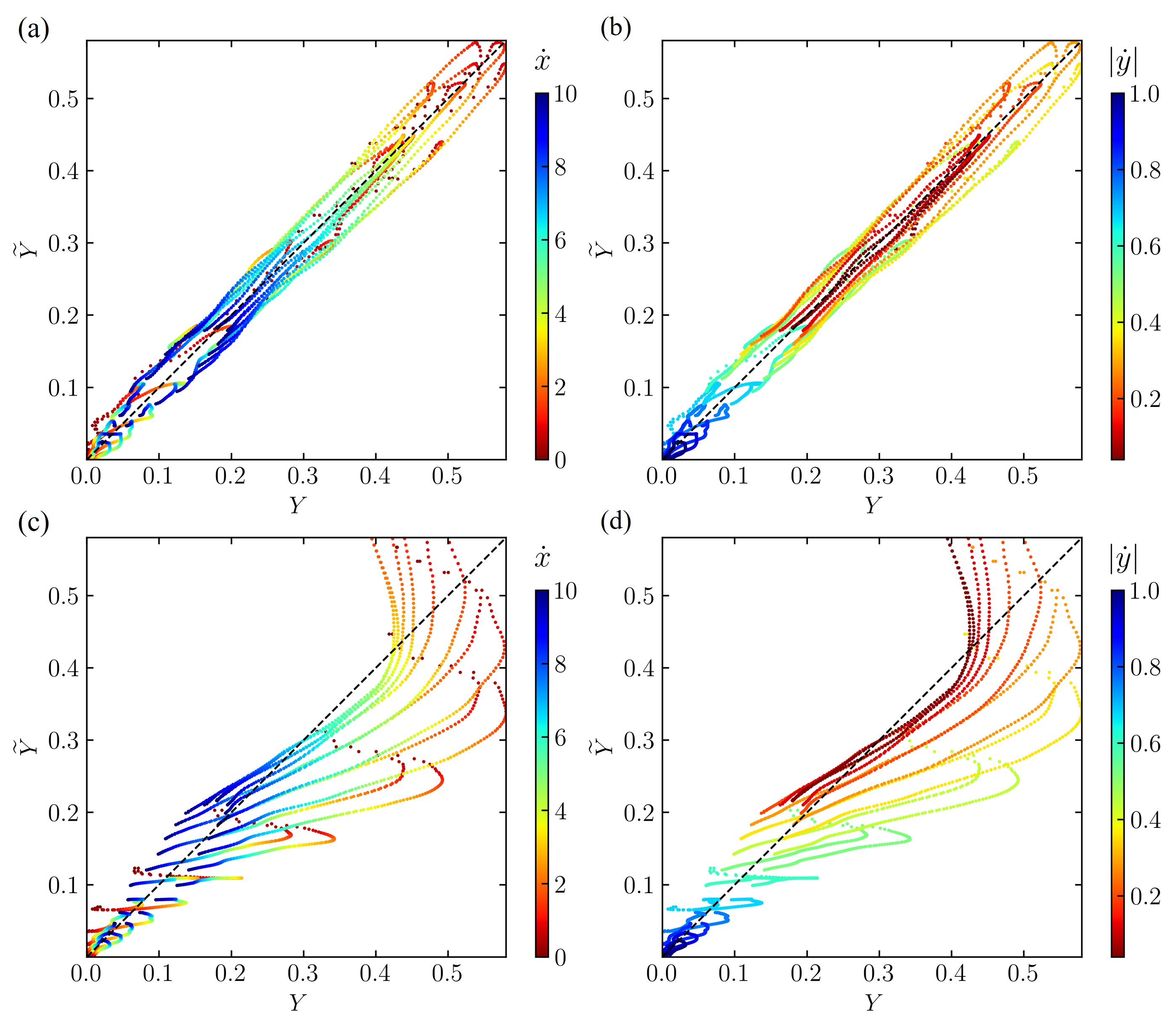}
    \caption{Identity plots between the normalized velocity deficit from CFD $(Y)$ on the x-axis and the results predicted $(\widetilde{Y})$ on the y-axis by SR in (a) and (b), and the Bastankhah wake model in (c) and (d). Data points are colored based on their streamwise positions in (a) and (c), and spanwise positions in (b) and (d), respectively. The black dashed line denotes perfect match.}
    \label{fig:y_x}
\end{figure*}

We use statistical methods to further evaluate the performance of the expression found by SR as a data-driven approach. In identity plots Fig.~\ref{fig:y_x}, the true values of the normalized wake velocity deficit and the expression obtained from knowledge discovery are utilized to assess the model's accuracy. The closer the data points are to the black dashed line with a slope of one, the more accurate the expression predictions are. We found that the discovered expression can accurately predict wake velocity within various value ranges, a crucial aspect for wind resource assessment. Figure~\ref{fig:y_x}(a) illustrates that errors are slightly greater in certain areas of the near-wake region ($x<2D$) and distant regions of the far-wake region ($x>8D$). In Fig.~\ref{fig:y_x}(b), the points where $|\dot{y}| \approx 0$ are almost all on the diagonal line, indicating that the predictions at the wake center of the full wake are extremely accurate. The errors of the velocity deficit away from the wake center are also small because the DG expression describes the wake deficit near zero at these positions, consistent with the fact that there is almost no interference far from the wake center. However, the Bastankhah wake model performs poorly when predicting flows with large velocity deficit, which occur in the mid and root of the blade in the near wake. Fig.~\ref{fig:y_x}(d) indicates that in the spanwise direction, the wake model tends to overestimate the wake deficit at the wake center first and then underestimate that away from the wake center until $|\dot{y}|\gtrsim 0.6$.

\begin{figure*}
    \centering
    \includegraphics[width=\linewidth]{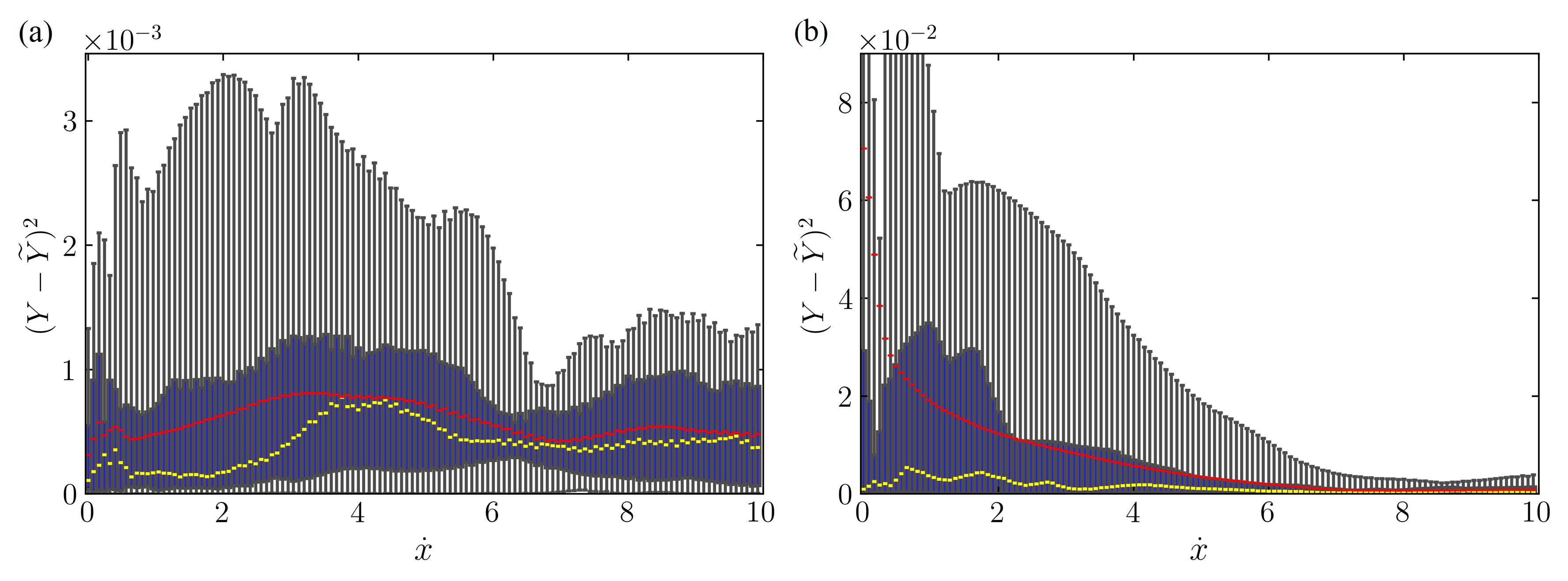}
    \caption{Prediction errors along the streamwise direction of the velocity deficit from (a) the SR expression and (b) the Bastankhah wake model. The upper and lower horizontal lines indicate maximum and minimum values, and the blue boxes represent interquartile ranges. The red line and the yellow line indicate mean and median respectively.}
    \label{fig:box}
\end{figure*}

Fig.~\ref{fig:box} illustrates the errors in the streamwise direction of velocity deficit predictions obtained from the SR expression and the Bastankhah wake model. We find that the data-driven method maintains relatively consistent average errors across different positions along the streamwise direction. The maximum average error occurs in the range of $3D < x < 5D$, although this value is significantly smaller compared to the results from the wake model. The proximity of mean and median suggests that the error distribution is roughly uniform in the spanwise direction, indicating that the SR expression accurately describes the wake deficit at both the wake center and the wake edge. Thus, we believe that the expression derived through knowledge discovery can effectively predict the average velocity field of the full wind-turbine wake. In contrast, the prediction accuracy of the Bastankhah wake model in the near wake is inadequate, with a difference of one order of magnitude compared to the SR expression. In the near wake, the mean is significantly larger than the median and even surpasses the upper quartile, indicating a few spanwise positions where the predictions are quite inaccurate. However, due to the applicability of the Gaussian profile in the far wake, the Bastankhah wake model performs well.

\begin{figure}
    \centering
    \includegraphics[width=\linewidth]{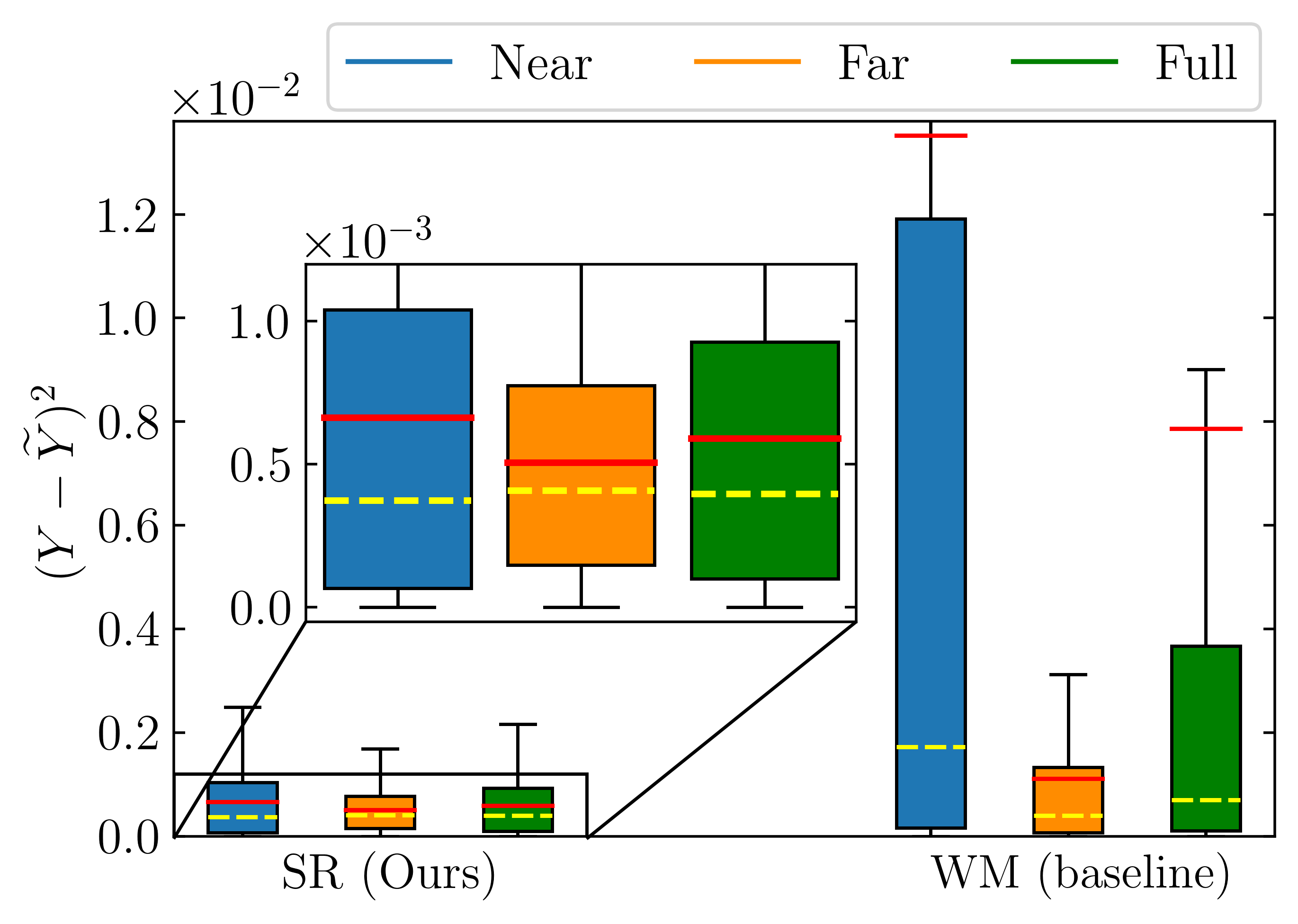}
    \caption{Comparison of prediction errors of the velocity deficit in the near-wake region, far-wake region and the full wake from the SR expression and the Bastankhah wake model. Mean and median are denoted by the red line and the yellow line respectively.}
    \label{fig:near}
\end{figure}

Traditional wake models have always struggled in accurately predicting the velocity deficit in the near wake due to their reliance on simplified conservation equations and the assumption of wake deficit shape, which have limited applicability in this region. Therefore, we respectively compared the prediction errors of the velocity deficit in the near-wake region, far-wake region, and the full wake from the SR expression and the Bastankhah wake model, as shown in Figure \ref{fig:near}. Typically, the boundaries between the near-wake and far-wake regions are considered as where the velocity deficit transformed into a Gaussian distribution. This boundary $X_{cr}$ here is estimated by the empirical formula proposed by Bastankhah \textit{et al.}~\cite{Bastankhah2016}:
\begin{equation}
\frac{X_{cr}}{D}=\frac{1+\sqrt{1-C_{\mathrm{T}}}}{\sqrt{2} \left(4 \alpha T I_{\mathrm{u}}+2 \beta\left(1-\sqrt{1-C_{\mathrm{T}}}\right)\right)},
\end{equation}
where $\alpha=0.58$ and $\beta=0.077$. We find that the errors of the expression obtained from SR in the near wake are significantly smaller than those of the wake model. Consequently, the full-wake prediction errors of the former are also smaller than the latter. The errors of the SR expression remain consistent among the three regions, indicating their robustness in capturing the mean flow at various positions in the full wake. Conversely, the wake model introduces an average error that is much larger than the upper quartile in the full-wake prediction, suggesting that large errors at a few positions adversely impact the overall performance of this method. Such significant errors can have detrimental effects on average evaluation in engineering applications. When considering the far wake, the Bastankhah wake model exhibits only slightly worse performance when compared to the data-driven method.

\section{Validation}
\label{sec:validation}
To comprehensively validate the SR expression derived from our study, we employ both experimental data~\cite{Keane2021,Schreiber2020} and high-fidelity numerical simulations~\cite{Soesanto2023}. This validation process encompasses different real-world scenarios to ensure the robustness and accuracy of the SR expression.

\begin{figure*}
    \centering
    \includegraphics[width=0.75\linewidth]{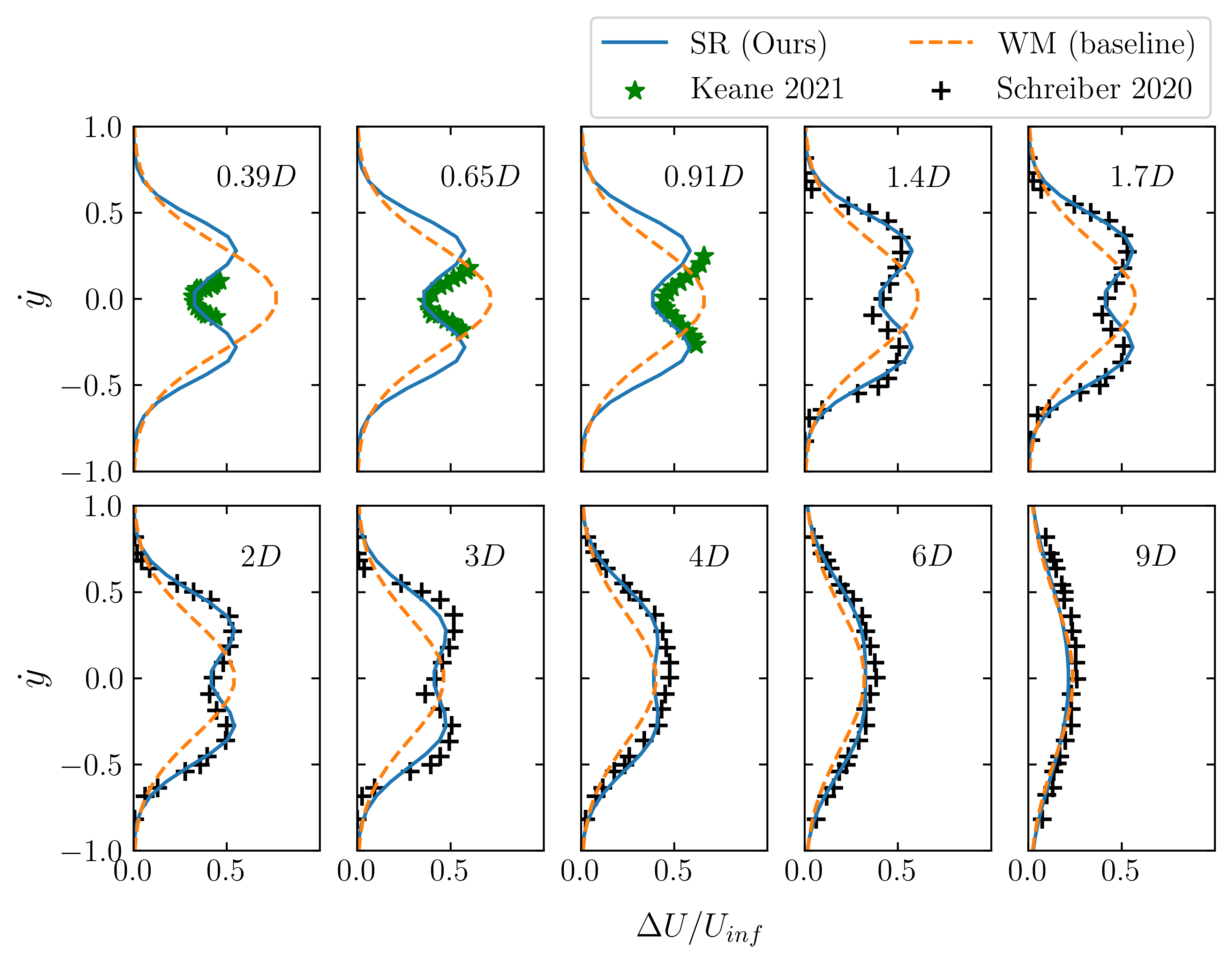}
    \caption{Comparison of velocity deficit profiles at several stations from the SR expression, the Bastankhah wake model and the field measurements as well as wind tunnel experiment data.}
    \label{fig:val1}
\end{figure*}

We utilize experimental data to validate the performance of the SR expression, particularly in the near wake. In the study of keane \textit{et al.}~\cite{Keane2021}, quite a few field measurements in the near wake, including around the turbine rotor, are available. The authors employed a laser radar installed on the wind turbine nacelle to scan the flow field and obtain the radial line-of-sight velocity. However, they did not provide any information regarding turbulence intensity in their paper. In another study \cite{Gao2022}, the incoming turbulence velocity in Keane's work was estimated to be $8\%$. Schreiber \textit{et al.}~\cite{Schreiber2020} measured the wake velocity at seven positions downstream of the turbine using hot-wire probes in a wind tunnel, and the turbulence intensity at hub height was found to be approximately $5\%$ using a pitot tube. Fig.~\ref{fig:val1} compares the wake velocity deficit predicted by the SR expression with the experimental results. The expression aligns well with the experimental results, from the far wake to the near wake and even near the turbine rotor. However, for field experiment data, the SR expression slightly underestimates the velocity deficit at $x = 0.91D$. This discrepancy is due to the fact that, at this location, the wake deficit near the turbine rotor increases along the streamwise direction, and the higher turbulence intensity causes the wake to develop faster. On the other hand, the prediction results demonstrate high accuracy for wind tunnel experiments, including in the near wake. However, analysis of the wind tunnel data indicates that the wake transitions from a DG to a Gaussian distribution earlier than predicted.

\begin{figure}
    \centering
    \includegraphics[width=\linewidth]{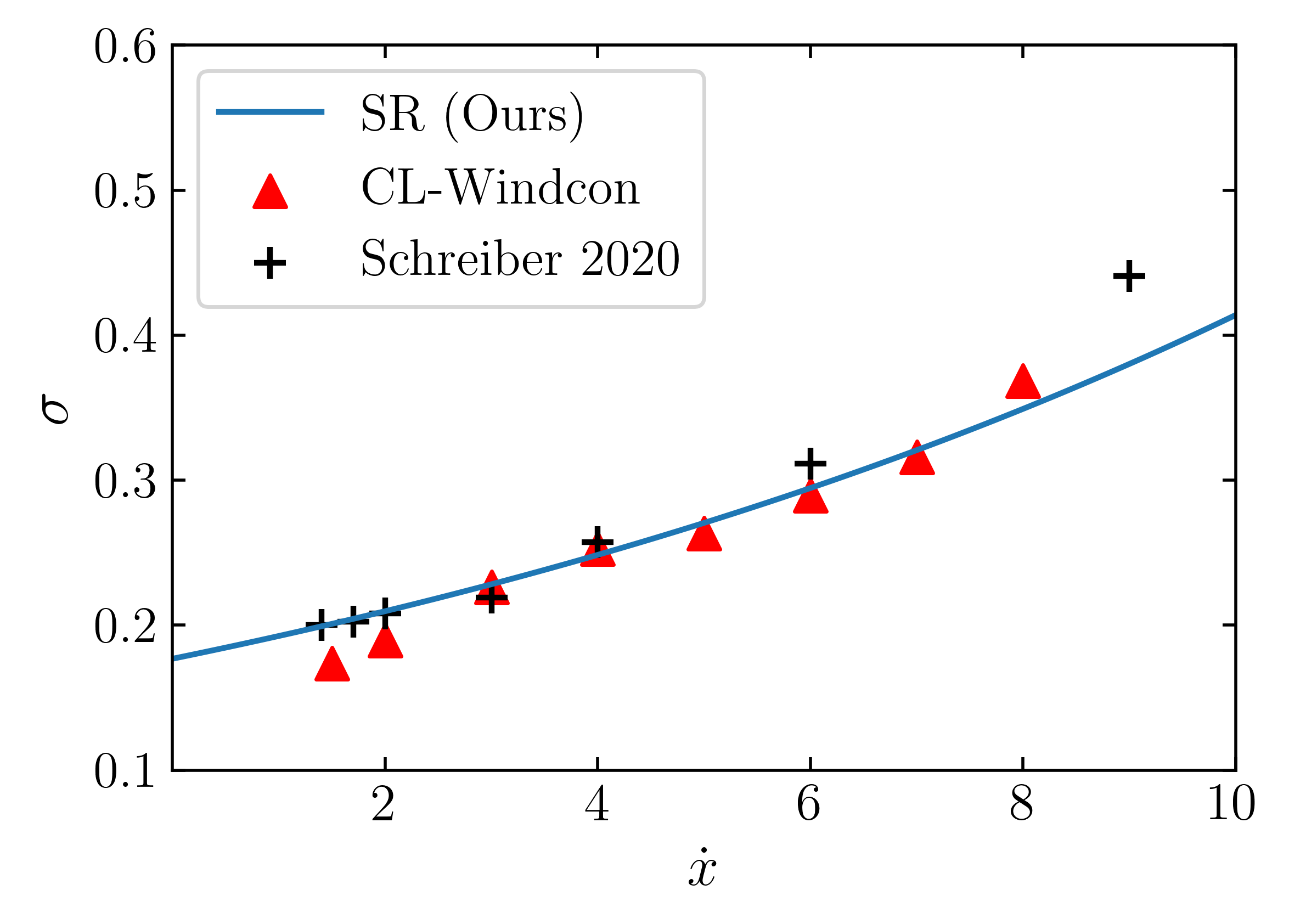}
    \caption{The streamwise progress of the parameter standard deviation $\sigma$ from the SR expression, the wind tunnel experiment and the high-fidelity numerical simulations.}
    \label{fig:val2}
\end{figure}

In addition to describing the distribution of velocity deficit at each downstream location, another critical assumption of traditional wake models is that the wake expands downstream in a linear way. Studying the axial behavior of the parameter standard deviation $\sigma$ is therefore helpful for wake modeling. Classical theoretical studies believed that the wake expansion correlates with the cubic root of streamwise positions, meaning that $\sigma$ increases with downstream distance in proportion to $x^{1/3}$ \cite{Tennekes1972AFC,Keane2016}. However, since this theory did not account for the impact of ambient turbulence on wake recovery, subsequent studies have approximated the wake width increasing linearly with $x$ \cite{Bas}. In Fig.~\ref{fig:val2}, we present the variation of the parameter standard deviation $\sigma$ with downstream distance. Soesanto \textit{et al.}~\cite{Soesanto2023} employed a DG fit to investigate the standard deviation $\sigma$ at different downstream locations using wake data from CL-Windcon project. The project utilized LES with higher resolution to simulate wind turbine wakes, employing the high-fidelity actuator line method for indirect modeling of the rotor. The turbulence intensity at the hub height is recorded as $4.1\%$. We observed that the SR expression for standard deviation $\sigma$ of the DG profile is reasonable. For wind-turbine wake fields with relatively low turbulence intensity, such as in this study, the wake expansion is not uniform but rather accelerates, especially in the further downstream region $(x > 8D)$. This phenomenon of wake recovery at low turbulence intensity has also been observed in previous studies \cite{Bas}. Hence, we believe that knowledge discovery methods not only extract expressions from data that more precisely describe physical phenomena compared to those derived from simplified physical laws but also provide interpretable expressions that facilitate deeper comprehension and exploration of the underlying physical mechanisms.

\section{Conclusions}
\label{sec:conclu}
In modern wind farms, turbines are increasingly placed in closer proximity, emphasizing the need for a comprehensive understanding of the aerodynamic properties of wind turbine wakes for practical engineering purposes. This study has employed a knowledge discovery approach to model a full wind-turbine wake, including the near-wake region. The genetic SR algorithm was used to extract an interpretable expression as a previously unavailable insight, capable of describing the mean wake velocity deficit. By incorporating the DG profile of the velocity deficit as domain knowledge into the SR algorithm and then formulate a hierarchical equation structure, the search space was reduced, resulting in a succinct, physically informative and robust wake model.

Traditional wake models derived from simplified conservation equations and assumptions have not adequately addressed turbine wakes in the near-wake region, especially those in proximity to the rotor. However, the SR expression selected based on fitness and scoring system can accurately represent the velocity deficit of a full wake. Unlike previous wake models designed for the far-wake region, the SR expression can predict the velocity deficit at any location in the wake field with high precision and stability. The effectiveness and accuracy of this knowledge-discovered wake model are further validated through experiments and high-fidelity numerical simulations. Furthermore, the SR expression can accurately describe the following wake performances in a simple mathematical form: (i) the wake velocity deficit increases and then decreases along the streamwise direction, (ii) the peak locations of the velocity deficit in the DG function remain relatively fixed in the spanwise direction, and (iii) the wake expands faster in the farther downstream far-wake region under relatively low turbulence intensity. The findings illustrate that knowledge discovery methods have the capacity to surpass the constraints imposed by ideal assumptions in theoretical derivations. This enables the construction of more accurate models that align closely with observed data in real-world scenarios, fostering enhanced understanding and exploration of the underlying physical mechanisms.

In general, both experimental and simulation data regarding specific physical process offer a wealth of information. However, it is crucial to distill the most valuable insights in an interpretable manner, while ensuring adherence to physical laws. This study illustrates that the proposed domain knowledge-driven genetic SR framework is a viable and effective solution to this challenge. It contributes to the advancement of human cognitive boundaries and furnishes empirical support for the application of interpretable data-driven techniques in wind farm research. It is noteworthy that the turbine wake under investigation occurs in the turbulent ABL, characterized by a relatively low turbulence intensity, and the potential impact of varying turbulence intensities on the model has not been taken into account. Additionally, velocity shear due to ground effects requires further consideration when applying wake models in the vertical direction. While the PySR library is mature and user-friendly, its reliance on the genetic algorithm may result in inefficiencies. Hence, future research endeavors will explore the utilization of reinforcement learning-based methods, such as DISCOVER \cite{DISCOVER}, to further enhance the efficiency. Subsequent research should prioritize addressing these limitations within the proposed modeling framework.

\nocite{*}

\begin{acknowledgments}
This work was supported by the National Natural Science
Foundation of China (Grant No. 62106116),  China Meteorological Administration under Grant QBZ202316, Natural Science Foundation of Ningbo of China (No. 2023J027), as well as by the High Performance Computing Centers at Eastern Institute of Technology, Ningbo, and Ningbo Institute of Digital Twin.
\end{acknowledgments}

\section*{Data Availability}
The data supporting the findings of this study are available upon reasonable request.

\appendix
\section{Evolution operations of mutation and crossover}
\label{app:evol}
\begin{figure*}
    \centering
    \includegraphics[width=0.85\linewidth]{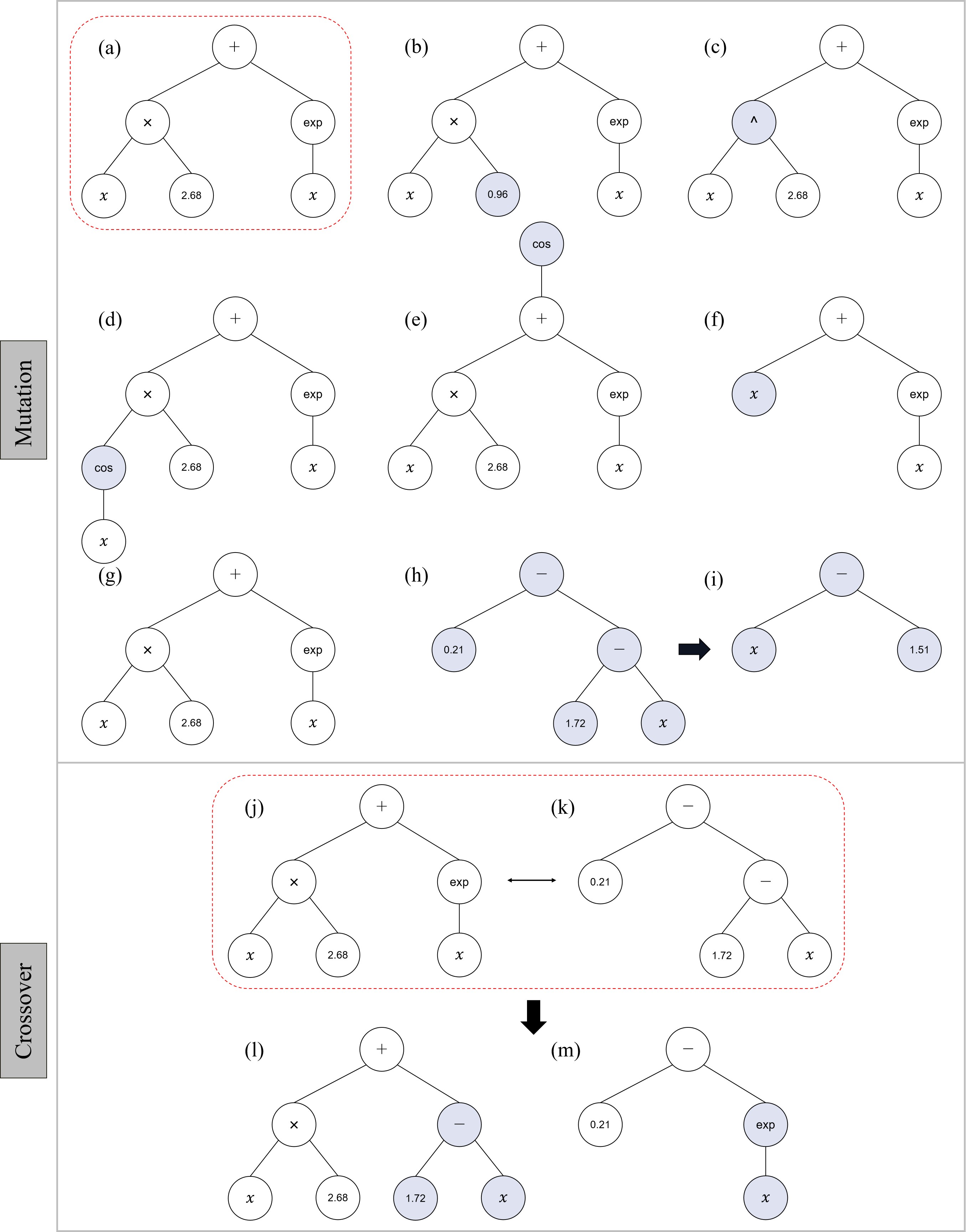}
    \caption{Evolution operations of mutation and crossover in the genetic algorithm. Expression trees enclosed within red boxes represent instances chosen via tournament selection for evolution. Beginning with the original individual (a), diverse mutation strategies are executed optionally, encompassing (b) constant mutation, (c) operator mutation, (d) node insertion, (e) node addition, (f) subtree deletion, (g) no operation, (h) tree replacement and (i) expression simplification. The subsequent generation of individuals (l) and (m) results from the crossing over of parental individuals (j) and (k).}
    \label{fig:evlution}
\end{figure*}

Genetic algorithms can generate an extensive pool of variants by creating unlimited combinations of genes, thereby significantly expanding the search for potential expressions. In this study, the genetic algorithm utilized eight mutation options along with crossover mechanisms to realize evolution, as shown in Fig.~\ref{fig:evlution}.

\section{Expressions discovered by SR}
\label{app:exp}
We simultaneously perform SR on three parameters: amplitude $a$, mean $\mu$, and standard deviation $\sigma$, each with a maximum complexity of $c_{a,max}=c_{\mu,max}=c_{\sigma,max}=20$. This process results in three distinct sets of Pareto fronts, as illustrated in Tab.~\ref{tab:a}, Tab.~\ref{tab:mean} and Tab.~\ref{tab:std} respectively. In the Pareto front, expressions with the highest accuracy at each level of complexity are displayed. These expressions must also demonstrate a decreasing error as complexity increases. Expressions that do not meet this criterion, therefore are excluded from the Pareto front.

\begin{table}[]
\raggedright
  \caption{Expressions for amplification $a$.}
  \scalebox{1}{
\begin{tabular}{p{0.8cm} p{7.2cm}}
\toprule
c  & Expressions                                                                                                               \\ \midrule
1  & $0.34$                                                                                                                    \\
3  & $0.78^{\dot{x}}$                                                                                                          \\
4  & $\cos{\left(1.0^{\dot{x}} \right)}$                                                                                       \\
5  & $0.59 - 0.052 \dot{x}$                                                                                                    \\
6  & $e^{- 0.52 \times 1.2^{\dot{x}}}$                                                                                         \\
7  & $0.21 e^{\cos{\left(0.22 \dot{x} \right)}}$                                                                               \\
8  & $0.80^{\dot{x}} - 0.58 e^{- \dot{x}}$                                                                                     \\
9  & $0.83^{\dot{x}} + \frac{0.45}{- \dot{x} - 0.77}$                                                                          \\
10 & $0.83^{\dot{x}} + \frac{0.45}{- \dot{x} - 0.77}$                                                                          \\
11 & $0.85^{\dot{x}} - 0.046 + \frac{0.39}{- \dot{x} - 0.73}$                                                                  \\
13 & $1.0 \times 0.85^{\dot{x}} - 0.045 + \frac{0.41}{- \dot{x} - 0.76}$                                                                                                                             \\
15 & $0.81^{\dot{x}} e^{- \frac{0.72}{2.1 \dot{x} + \cos{\left(\dot{x} + 0.59 \right)}}}$                                                                                                             \\
16 & $0.81^{\dot{x}} e^{- \frac{0.72}{2.1 \dot{x} + \cos{\left(\dot{x} + 0.61 \right)}}}$                                                                                                             \\
17 & $1.0 \times 0.80^{\dot{x}} e^{- \frac{0.78}{2.1 \dot{x} + \cos{\left(\dot{x} + 0.52 \right)}}}$       \\
19 & $1.0 \times 0.80^{\dot{x}} e^{- \frac{0.78}{2.1 \dot{x}^{1.0} + \cos{\left(\dot{x} + 0.49 \right)}}}$                    \\ \bottomrule
\end{tabular}}
\label{tab:a}
\end{table}

\begin{table}[]
\raggedright
  \caption{Expressions for mean $\mu$.}
  \scalebox{1}{
\begin{tabular}{p{0.8cm} p{7.2cm}}
\toprule
c  & Expressions                                                                                                         \\ \midrule
1  & $0.29$                                                                                                              \\
5  & $- \frac{86.}{- \dot{x} - 2.9 \times 10^{2}}$                                                                       \\
6  & $0.29 + 0.017 e^{- \dot{x}}$                                                                                        \\
7  & $\frac{0.28}{\cos{\left(\dot{x} e^{- \dot{x}} \right)}}$                                                            \\
8  & $0.011 \cos{\left(\dot{x} - 1.6 \right)} + 0.29$                                                                    \\
9  & $1.3 - e^{- 0.011 \cos{\left(\dot{x} - 1.6 \right)}}$                                                               \\
10 & $0.021 \cos{\left(1.3^{\dot{x}} - \dot{x} \right)} + 0.28$                                                          \\
11 & $0.013 e^{\cos{\left(1.3^{\dot{x}} - \dot{x} \right)}} + 0.27$                                                      \\
12 & $0.018 \cos{\left(- 1.3^{\dot{x}} + \dot{x} + 0.25 \right)} + 0.28$                                                 \\
14 & $0.018 \cos{\left(- 1.3^{\dot{x}} + 1.0 \dot{x} + 0.26 \right)} + 0.28$                                             \\
15 & $0.0037 \cos{\left(\dot{x} \right)} + 0.023 \cos{\left(1.3^{\dot{x}} - \dot{x} \right)} + 0.28$                     \\
16 & $0.0033 e^{\cos{\left(\dot{x} \right)}} + 0.023 \cos{\left(1.3^{\dot{x}} - \dot{x} \right)} + 0.27$                 \\
18 & $0.0032 e^{\cos{\left(\dot{x} \right)}} + 0.016 \cos{\left(1.3 \times 1.3^{\dot{x}} - 1.3 \dot{x} \right)} + 0.28$  \\
20 & $0.0028 e^{0.081 \dot{x} + \cos{\left(\dot{x} \right)}} + 0.025 \cos{\left(1.3^{\dot{x}} - \dot{x} \right)} + 0.27$ \\ \bottomrule
\end{tabular}}
\label{tab:mean}
\end{table}

\begin{table}[]
\raggedright
  \caption{Expressions for standard deviation $\sigma$.}
  \scalebox{1}{
\begin{tabular}{p{0.8cm} p{7.2cm}}
\toprule
c  & Expressions                                                                                          \\ \midrule
1  & $0.28$                                                                                               \\
5  & $0.18 \times 1.1^{\dot{x}}$                                                                          \\
6  & $0.18 e^{0.085 \dot{x}}$                                                                             \\
7  & $0.20 \times 1.1^{\dot{x}} - 0.027$                                                                  \\
8  & $0.18 \left(e^{\dot{x}} - 0.23\right)^{0.084}$                                                       \\
9  & $0.18 \left(1.1 - (3.0 \times 10^{-5})^{\dot{x}}\right)^{\dot{x}}$                                     \\
10 & $0.18 \left(0.00088 \cos{\left(\dot{x} \right)} + 1.1\right)^{\dot{x}}$                              \\
11 & $0.18 \left(1.1 - 0.0010 e^{- \cos{\left(\dot{x} \right)}}\right)^{\dot{x}}$                         \\
12 & $0.18 \left(0.0025 \cos{\left(1.7 \dot{x} \right)} + 1.1\right)^{\dot{x}}$                           \\
13 & $0.18 \left(0.0022 e^{\cos{\left(1.7 \dot{x} \right)}} + 1.1\right)^{\dot{x}}$                       \\
15 & $0.18 \left(0.0022 e^{\cos{\left(0.12 \dot{x}^{2} \right)}} + 1.1\right)^{\dot{x}}$                  \\
17 & $0.18 \left(0.0022 e^{\cos{\left(0.12 \dot{x}^{2} - 0.20 \right)}} + 1.1\right)^{\dot{x}}$           \\
19 & $0.18 \left(0.0022 e^{\cos{\left(0.68^{\dot{x}} - 0.12 \dot{x}^{2} \right)}} + 1.1\right)^{\dot{x}}$ \\ \bottomrule
\end{tabular}}
\label{tab:std}
\end{table}

\section{The Bastankhah wake model}
\label{app:Bas}

The wake velocity deficit in the far-wake region has been shown by Bastankhah \textit{et al.}~\cite{Bas} to adhere to a self-similar Gaussian distribution. The Gaussian analytical wake model they proposed is expressed as:
\begin{equation}
\frac{\Delta U}{U_{inf}}=\left(1-\sqrt{1-\frac{C_{T}}{8\left(\sigma/D\right)^{2}}}\right) \times \exp \left(-\frac{r^2}{2\sigma^2} \right).
\end{equation}
The assumption of linear wake expansion is utilized, as indicated by the standard deviation $\sigma$. The definition of $\sigma / D=k_b x / D+\varepsilon$ is employed, where $\varepsilon=0.25\sqrt{\beta}$ ensures consistency with the predicted rate of the initial mass flow deficit from Frandsen \textit{et al.}~\cite{Frandsen}. The researchers disregarded the streamwise distance necessary for pressure recovery, thereby positing that the initial wake area and rotor area share the relationship $A_w(x=0)=\beta A$. The value of $\beta$ is determined by:
\begin{equation}
\beta=\frac{1+\sqrt{1-C_T}}{2\sqrt{1-C_T}},
\end{equation}
guaranteeing a solution for all thrust coefficient ($C_T$) values ranging from 0 to 1.

\bibliography{Ref}

\end{document}